%% file: main.tex
\documentclass{article}
\usepackage{spconf,amsmath,graphicx}
\usepackage{multirow}
\usepackage{booktabs}
\usepackage{cite}
\usepackage{xcolor}
\usepackage{verbatim}
\usepackage{amsfonts}

\title{Acoustic-aware Non-autoregressive Spell Correction with Mask Sample Decoding}
\name{Ruchao Fan$^1$\sthanks{Work done during an internship at Microsoft.}, Guoli Ye$^2$, Yashesh Gaur$^2$, Jinyu Li$^2$}
\address{$^1$University of California, Los Angeles, $^2$Microsoft Corporation}
\begin{document}
\ninept
\maketitle
\begin{abstract}
Masked language model (MLM) has been widely used for understanding tasks, e.g. BERT. Recently, MLM has also been used for generation tasks. The most popular one in speech is using Mask-CTC for non-autoregressive speech recognition. In this paper, we take one step further, and explore the possibility of using MLM as a non-autoregressive spell correction (SC) model for transformer-transducer (TT), denoted as MLM-SC. Our initial experiments show that MLM-SC provides no improvements on Librispeech data. The problem might be the choice of modeling units (word pieces) and the inaccuracy of the TT confidence scores for English data. To solve the problem, we propose a mask sample decoding (MS-decode) method where the masked tokens can have the choice of being masked or not to compensate for the inaccuracy. As a result, we reduce the WER of a streaming TT from 7.6\% to 6.5\% on the Librispeech test-other data and the CER from 7.3\% to 6.1\% on the Aishell test data, respectively.

\end{abstract}
\begin{keywords}
End-to-end ASR, Transformer-Transducer, Non-autoregressive Spell Correction, Masked Language Model
\end{keywords}

\input{Tex/intro}
\input{Tex/method}

\input{Tex/exp_setup}

\input{Tex/results}
\input{Tex/conclusion}

\vfill\pagebreak

\ninept
\bibliographystyle{IEEEbib}
\bibliography{strings,refs}

\end{document}

%% file: Tex/intro.tex
\section{Introduction}
\label{sec:intro}

End-to-end (E2E) speech modeling has gained significant success in recent years since its ability to use a single model to directly optimize the objective function of an ASR \cite{ChiuSWPNCKWRGJL18,HeSPMAZRKWPLBSL19,ZhangLSTMKK20,Li0G0Z020,MoritzHR20, li2022recent}. Among the E2E models, the neural transducer models such as RNN-T \cite{Graves-RNNSeqTransduction, HeSPMAZRKWPLBSL19, Li2019improving} and Transformer-Transducer (TT) \cite{yeh2019transformer, ZhangLSTMKK20, chen2021developing} are the most popular ones because they are naturally streaming. However, the streaming is built on a causal encoder where at the current step the model observes only the historical information and has no access the future information. 
To further improve the recognition accuracy of the neural transducer model, researchers proposed two-pass models\cite{SainathPRHPLVLS19}, deliberation models \cite{HuPSS21,HuSPP20,HuSNPS22,YeMLG22}, and spell correction models \cite{GuoSW19}.  

In two-pass models, an LAS decoder \cite{ChanJLV16} is added to the shared encoder to learn together with the RNN-T decoder. The LAS decoder is essentially an autoregressive generator. Using the LAS decoder for second-pass will take too much computational cost during inference. Two-pass models have a rescoring mode, which could be fast for the 2nd-pass. However, it selects only the best candidates among the n-best list of the first-pass results, and cannot correct error tokens in an utterance. In deliberation models, the LAS decoder accepts two sequences as input: 1) the first-pass results of the transducer, and 2) an additional non-causal encoder. The non-causal encoder provides global acoustic information for the LAS decoder. The deliberation model has the ability to correct error tokens. However, the LAS decoder is still an autoregressive generator. Using such a model for 2nd-pass causes high latency and is not computationally efficient. The current spell correction models also adopt an autoregressive decoder. As a consequence, in order to accelerate the the second-pass spell correction, using non-autoregressive models on top of the neural transducer might be the most appropriate solution. 

Non-autoregressive speech processing is first used in \cite{ChenWVZD21}. After that, many more non-autoregressive methods are proposed\cite{TianYTBZW20,FanCC021,FanC00A21,GaoPara21,NozakiK21,DengYWHCZ22,ChanSH0J20}. Among the methods, there are two that are appropriate to achieve non-autoregressive spell correction. One is Align-Refine\cite{ChiSK21}. The other is Mask-CTC\cite{Higuchi0COK20}. For Align-Refine, researchers use the alignment from the neural transducer as decoder input\cite{WangHS22a}, and apply CTC loss function to optimize the decoder. Since alignment also exists in CTC output space, the alignment from the CTC output can again be input to the decoder such that the decoder can iteratively refine the alignment. However, multiple decoding iterations in Align-refine are not efficient enough. In this paper, we propose an acoustic-aware masked language model for non-autoregressive spell correction. Similar to Mask-CTC, we first train an acoustic-aware masked language model decoder, but with the parameters in the shared encoder fixed. During inference, those transducer outputs with low confidence scores will be masked out. The acoustic-aware masked language model then corrects the low confidence tokens. Since we use masked language model as a spell correction model, we denote the method as MLM-SC. In our initial experiments, MLM-SC behaves well on the Aishell1 data\cite{BuDNWZ17}. However, it does not work well for the Librispeech data\cite{PanayotovCPK15}. The reason might be that the confidence scores from transducer output for the Librispeech data are not accurate as the Aishell data, which is a simpler task. Furthermore, English uses sub-word as modeling units. Correcting one sub-word does not equal the improvement of WER. However, Mandarin directly uses tokens as the modeling units. Correcting one token theoretically improves the CER on Aishell1.

To reduce this gap, we further propose a mask sample (MS) decoding method to make MLM-SC effective for English tasks. In mask sample decoding, when a token is masked because of low confidence scores, we give it a chance to be not masked. In other words, we sample between ``mask'' and ``not mask'' for this token. With multiple low confidence tokens, we can sample multiple mask solutions and input them to the MLM-SC. Then, multiple corrected utterances are generated. We use an external language model to select the best one as the final result. The MS decoding can compensate for the inaccuracy of the transducer confidence scores. Finally, MLM-SC achieves impressive improvements on the Librispeech data with the proposed MS decoding. One recent work \cite{Futami22narsc} uses phone2word conversion masked language model to achieve non-autoregressive spell correction. However, it does not perform well on English tasks.

The remainder of this paper is organized as follows. Section \ref{sec:methods} introduces the proposed acoustic-aware masked language model and mask sample decoding method. Experimental setups are described in Section \ref{sec:exp_setup}. Results are shown and discussed in Section \ref{sec:results}. We conclude the paper in Section \ref{sec:conclusion}.

%% file: Tex/method.tex
\section{Proposed Methods}
\label{sec:methods}
In this section, we introduce the neural transducer model and describe how to use the proposed acoustic-aware masked language model for spell correction. We further introduce the proposed mask sample decoding algorithm.

\subsection{Transducer Model}
\label{ssec:transducer model}
The transducer model consists of an audio encoder, a prediction network, and a joint network, as shown in Figure \ref{fig:mlm-sc}. The encoder consumes the acoustic feature $x^t_1$ and generates acoustic representations $h_t$. Given the label sequence $y^{u-1}_1$, the prediction network generates label representations $g_u$. The outputs of audio encoder and the prediction network are combined in the joint network. The output layer converts the hidden embeddings to the probability distribution over the vocabulary. The forward computation of the transducer model
could be written as below:

\begin{align}
    h_t &= AudioEnc(x_1^t) \\
    g_u &= PredNet(y_1^{u-1}) \\
    z_{t,u} &= W_o * max(0, h_t + g_u) \\
    P(\hat{y}_{t+1}|x_1^t,y_1^u) &= softmax(z_{t,u})
\end{align}

Similar to CTC, a special blank symbol, $\phi$, is
added to the output vocabulary to represent a null token.  The objective function of the transducer model is to minimize the negative log probability over all possible alignments, which could be written as:
\begin{equation}
L_\text{transducer} = p(y|x) = -log \sum_{\alpha \in \beta ^{-1} (y)} p(\alpha | x)
\end{equation}
where $\beta$ is to convert the alignment $\alpha$ to
the +label sequence $y$ by removing the blank $\phi$.

In this paper, we select the Transformer-Transducer (TT)
model to generate the first pass results. In TT, we use the transformer for the audio encoder, the LSTM for the prediction network.

\subsection{Acoustic-aware Masked Language Model}
\label{ssec:mlm}
Masked language model (MLM) achieves language modelling by accepting sentences with regions of mask tokens as input. MLM's ability is to reconstruct the masked tokens given all unmasked tokens. In this paper, we propose to use acoustic-aware masked language model as a spell correction model (MLM-SC), where the model would consider acoustic information when correcting masked tokens. If we define MLM input as $Y=\{y_1,y_2,...,y_U\}$, and $Y_{mask}$, $Y_{obs}$ are the masked and unmasked regions respectively, then the loss function for the acoustic-aware MLM can be described as: 
\begin{equation}
L_{\text{MLM}} = \sum_{y_t\in Y_{mask}}logP(y_t|Y_{obs}, h_1^T)
\end{equation}

\begin{figure}[t]
\centering
\centerline{\includegraphics[width=0.49\textwidth,height=0.35\textwidth]{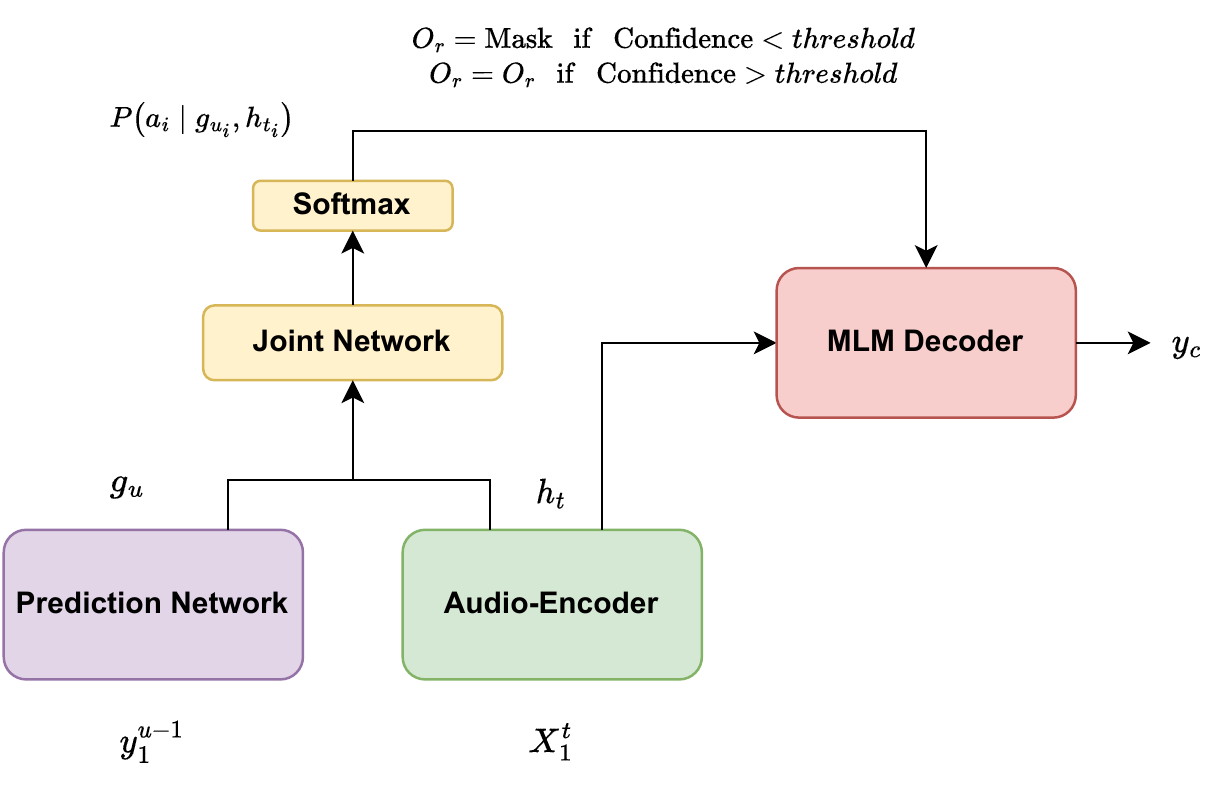}}
\caption{The schematic of using acoustic-aware masked language model as a spell correction model (MLM-SC). Confidence scores are obtained from transducer outputs.}
\label{fig:mlm-sc}
\end{figure}

MLM has a good property of correcting masked tokens given bidirectional contextual tokens. It can be excellent as a spell correction model, which compensates for the unidirectional language modeling of the prediction network. We plot how to use the acoustic-aware MLM as a spell correction model in Fig.\ref{fig:mlm-sc}. In Figure \ref{fig:mlm-sc}, there is a transducer model on the left. On the right side, a masked language model is integrated with the audio encoder. During training, the acoustic-aware MLM is optimized with the ground-truth as input. Each ground-truth is uniformly masked at most 40\% of the utterance length. The MLM reconstructs the masked tokens given unmasked tokens. During inference, the MLM accepts input from the audio encoder, as well as the transducer output with the low confidence tokens being masked. With the property of mask reconstruction, the acoustic-aware MLM can correct the tokens with low confidence scores theoretically. Hence, MLM can be used as a spell correction model. Since MLM is a naturally non-autoregressive model, using MLM-SC causes low latency for the system, which is suitable as a second-pass model.

\subsection{Mask Sample Decoding}
\label{ssec:msd}
In a Mandarin system, the modeling units are tokens. MLM-SC directly corrects tokens so that the character error rate (CER) will drop easily. While for an English system, people always use sub-word units as the modeling units. When MLM-SC corrects a sub-word unit, the word is not guaranteed to be corrected and the word error rate (WER) does not necessarily change. Furthermore, TT confidence scores of the English task may not be as accurate as the Mandarin task, which makes the correction of sub-word units harder. Note that the confidence score of a token is the forward probability when TT outputs the current token. To reduce the gap of using MLM-SC for Mandarin and English systems, we propose a mask sample (MS) decoding algorithm. In MS decoding, the mask solution is not determined even if the confidence scores and the threshold are determined. Instead, for each token with a low confidence score, the token can be masked or not. It is equivalent to sample between ``mask'' and ``not mask'' (True or False for the mask solution for the utterance). For example, if we have TT confidence scores [0.6, 0.8, 0.9, 0.9, 0.95, 0.8] and the threshold is 0.85. Without the MS decoding algorithm, the mask solution is [True, True, False, False, False, True]. However, we suppose the ground-truth mask solution is [False, True, False, True, False, False]. The difference is caused by the inaccuracy of the TT confidence scores. With MS decoding, the mask solutions can be:

\begin{itemize}
  \item $[False, True, False, False, False, True]$
  \item $[False, False, False, False, False, True]$
  \item $[False, True, False, False, False, False]$
  \item $[True, False, False, False, False, False]$
  \item $[True, True, False, False, False, True]$
  \item $[True, False, False, False, False, True]$
  \item $......$
\end{itemize}

High confidence tokens are kept unchanged as ``False''. The third mask solution in the above list is almost the same as the ground-truth mask solution. We expect such mask solutions can be well corrected by MLM-SC and this is why MS decoding can compensate for the inaccuracy of the TT confidence scores. Note that, the mask solution when MS decoding is not used is also included in the above list (the fifth item). Multiple mask solutions are sampled as the MLM-SC input (the above list). Later on, corresponding corrected utterances are generated. A language model will be used for selecting the best output among all MLM-SC outputs. When using the language model to select the best MLM-SC outputs, MLM-SC scores are also considered. The best ratio of the MLM-SC and language model scores from our experiments is 1.0:0.5.

%% file: Tex/exp_setup.tex
\section{Experimental Setup}
\label{sec:exp_setup}

The experiments are evaluated on the 960-hour LibriSpeech English corpus \cite{PanayotovCPK15} and the 178-hour Aishell1 Mandarin corpus \cite{BuDNWZ17}. All experiments use 80-dim Mel-filter bank features, computed every 10ms with a 25ms window. The sets of output labels consist of 5k word-pieces obtained by the Sentence-Piece method \cite{KudoR18} for Librispeech and 4230 Chinese characters for Aishell1, which are obtained from the training set. 

For Librispeech, the transformer transducer consists of 24 transformer encoder blocks with 512 attention dimensions and 2048 feed-forward dimensions, and 2 LSTM prediction networks. For Aishell1, the transformer transducer consists of 12 transformer encoder blocks with 256 attention dimensions and 1024 feed-forward dimensions, and 1 LSTM prediction network. All masked language models are 6 blocks transformer encoder with 512 attention dimenions and 2048 feed-forward dimensions. The MLM accepts audio encoder outputs as source information. We also add a non-causal encoder on top of the causal audio encoder. The non-casual encoder is an 4-block transformer encoder with 512 attention dimensions and 2048 feed-forward dimensions. The non-causal encoder uses global attention such that the MLM-SC can observe more acoustic information than when only the causal audio encoder is used. All models are optimized with the Adam optimizer. The scheduler is using a warmup for the first 25k and then linear decay the learning rate to zero for 400k steps (Librispeech). For Aishell, the warmup step is 15k and training steps are 150k. We first train a transducer model. When training the MLM-SC model, the parameters in the transducer model are frozen to keep the first-pass results unchanged. For mask sample decoding, 10 mask solutions are sampled with a threshold of 0.95 for each n-best result.

We use the language models in Espnet \cite{WatanabeHKHNUSH18}.
For Librispeech, a transformer-based language model is selected. For Aishell, we use a RNN language model.

%% file: Tex/results.tex
\section{Results and Discussion}
\label{sec:results}

\subsection{MLM-SC}
\label{ssec:MLM-SC}

\input{Tables/aishell_results}
We evaluate the acoustic-aware MLM-SC on the Aishell1 and Librispeech data first to see its effectiveness on spell correction. The results of MLM-SC on Aishell1 data are shown in Table \ref{tab:aishell}. From the table, we can observe MLM-SC achieves relative 11\% character error rate (CER) improvements compared to the TT-baseline, showing that MLM-SC is very effective for the Aishell1 data. We conduct one experiment that simulates insertion errors in MLM-SC training. However, training MLM-SC without simulating insertion errors is better. The reason is that there are few insertion errors in the Aishell data (See Figure \ref{fig:sid_errors}). We also try to obtain the results of using mask from reference (alignment between the reference and hypothesis). The number of 4.8\% is the lower bound CER of the system. 
\input{Tables/librispeech_results}

The results of the acoustic-aware MLM-SC on the Librispeech are presented in Table \ref{tab:libriresults}. We conduct three experiments for MLM-SC training: 1) masking word; 2) masking token; 3) masking token + simulate insertion errors. From the table, we can observe that the performance of masking token is better than that of masking word. Further, simulating the insertion errors brings slight improvements compared to no simulation of the insertion errors because the Librispeech data contain more insertion errors than the Aishell data (shown in Figure \ref{fig:sid_errors}). We finally choose the masking token + simulating insertion errors as the MLM-SC training strategy for the Librispeech data. However, the WER of MLM-SC is close to the TT-baseline, which indicates MLM-SC does not work for the Librispecch data. This is reasonable because when correcting a sub-word unit, the word may still be wrong. On the other hand, if we mask correctly using reference, the lower bound WER of MLM-SC is 6.6\% on the test-other data, indicating the inaccuracy of the TT confidence scores.

\input{Tables/ms_decode}
\subsection{Mask Sample Decoding}
\label{ssec:msdecoding}
To reduce the performance gap between the character-based and sub-word unit systems, mask sample (MS) decoding is proposed. The results of MS decoding are shown in Table \ref{tab:ms_decode}. In the table, the first three lines are the WERs of the TT-baseline and acoustic-aware MLM-SC. We can again see that there are no improvements when using MLM-SC on the Librispeech data. When we apply MS decoding, the WER on the Librispeech data drops significantly. First, when correcting the 1-best result, the WER drops from 7.6\% to 7.1\% on the test other data. The oracle WER is when we choose the best output (compared to reference) for each test sample without using a language model. We test the oracle WER to see the lower bound of the WER that MS-decoding can achieve. When correcting all 5-best results in the n-best list, using MS decoding can improve the WER of MLM-SC from 7.6\% to 6.7\% on the test other data. The results show that MS decoding is effective to compensate for the inaccuracy of TT confidence scores, and is helpful to reduce the gap between the Librispeech (English) and Aishell1 (Mandarin) data. 

Additionally, we apply SpecAugment \cite{ParkCZCZCL19}, and using a non-causal encoder on top of the causal audio encoder to further improve the performance. As a result, we can achieve a WER of 6.9\% on the Librispeech test-other data when correcting 1-best result,. When correcting all 5-best results in the n-best list, the WER is dropped to 6.5\%, which means we have a 14.5\% relative WER improvements compared to the TT-baseline.

\subsection{Overall Results}
\label{ssec:over_results}
\input{Tables/overall_results}

Finally, we aggregate all the results together and compare their WERs on the Librispeech data, CERs on the Aishell data, and real time factors (RTF). The results are presented in Table \ref{tab:overall_results}. We add the results of beam search decoding with a size of 50 for readers' convenience because when correcting 5-best results, 50 masked solutions are sampled. As we can see from the table, if we compare the results of MLM-SC + MS decoding with bm50 + rescore, the improvements of MLM-SC becomes smaller. However, the RTF for bm50 is around 1.6, which is not practical to be used as a 2-pass model. Hence, we compare the proposed methods to bm5 results with rescoring. The proposed acoustic-aware MLM-SC with MS-decode achieves 7.0\% relative WER improvements on the librispeech test other data, and 11.6\% relative CER improvements on the aishell test data, respectively. The RTF of MLM-SC is nearly the same as TT-baseline, which is expected, because of the non-autoregressive mechanism of MLM-SC.

\begin{figure}[t]
\centering
\centerline{\includegraphics[width=0.45\textwidth,height=0.30\textwidth]{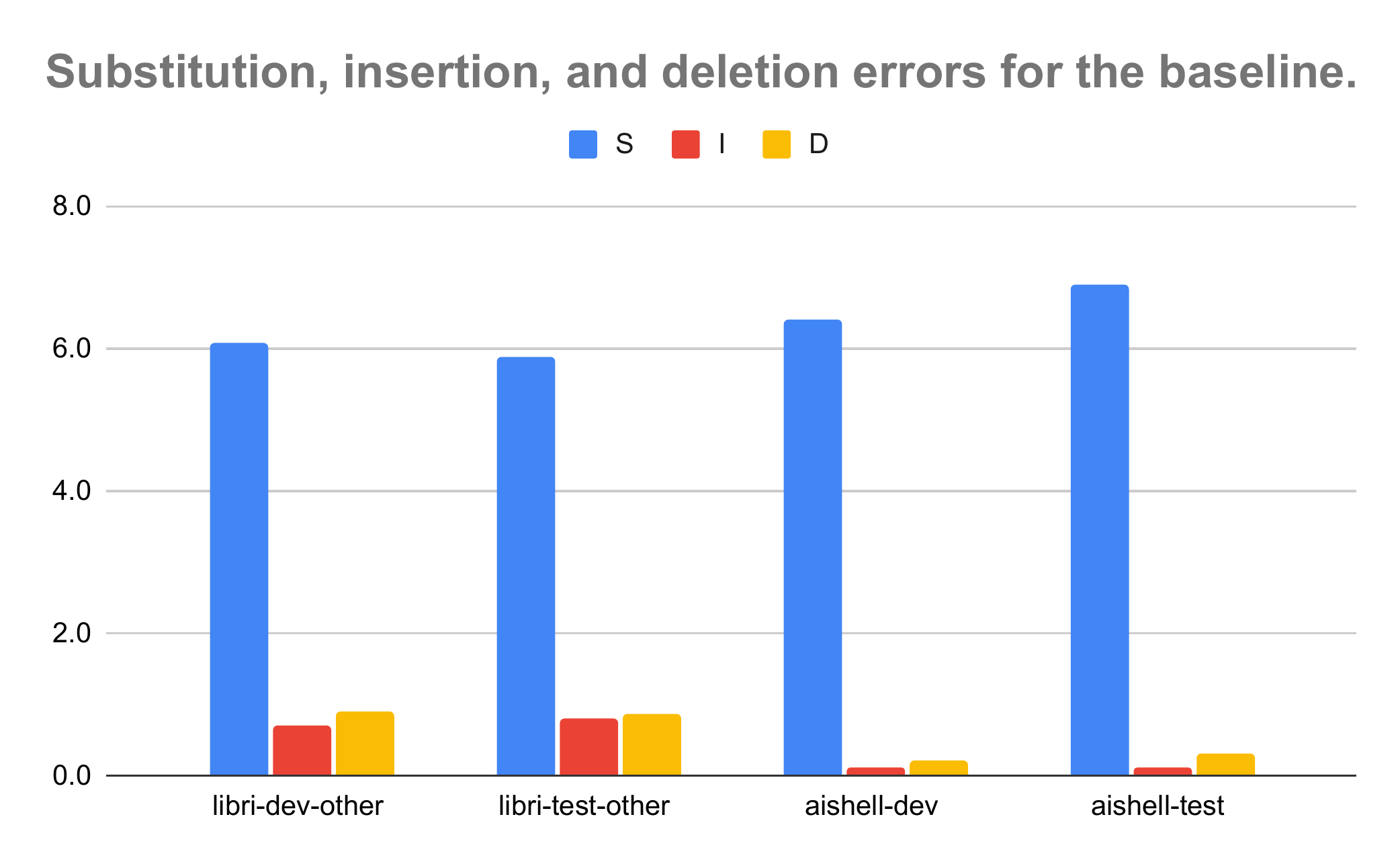}}
\caption{The substitution, insertion, and deletion errors on the Librispeech dev-other, test-other data, Aishell dev and test data, respectively. The evaluated models are the TT-baseline. ``S": Substitution error; ``I": Insertion error; ``D": Deletion error.}
\label{fig:sid_errors}
\end{figure}

One may argue that MLM-SC cannot solve deletion error problems. Hence, we analyse the substitution, insertion, and deletion errors for the Librispeech and Aishell1 data. All kinds of errors are shown in Figure \ref{fig:sid_errors}. We can see from the figure that most of the errors in the datasets are substitution errors. The summation of insertion and deletions errors are smaller than 5\% of the all errors for the Aishell data. With most of the errors being substitution errors, we infer that the proposed solution corrects most of the errors in the first pass results, which is the reason why MLM-SC with MS decoding is effective for both the Librispeech and Aishell data.

%% file: Tables/aishell_results.tex
\begin{table}[tp]
\caption{\small {Results of MLM-SC on the Aishell data. ``from reference" indicates the mask is obtained from alignment between the reference and hypothesis (lower bound of CER for the system). "sim. ins" represents simulating insertion errors in MLM-SC training. ``bm5" indicates using beam size of 5 during beam search decoding.}}
\normalsize
\vspace*{2mm}
\centering
\begin{tabular}{l l cc}
\hline
\multirow{2}{*}{} & \multirow{2}{*}{\shortstack{Decoding \\ Condition}} & \multicolumn{2}{c}{CER (\%)} \\
~ & ~ & dev &  test  \\
\hline
TT-baseline & bm5 & 6.7 & 7.3 \\
\hline
\multirow{2}{*}{mask token} & from TT score & 6.0 & 6.5 \\
 & from Reference & 4.4 & 4.8 \\
 + sim. ins. & from TT score & 6.4 & 7.0 \\
\hline
\end{tabular}
\label{tab:aishell}
\end{table}

%% file: Tables/librispeech_results.tex
\begin{table}[tp]
\caption{\small {Results of MLM-SC on the Librispeech data.}}
\footnotesize
\vspace*{2mm}
\centering
\begin{tabular}{l l cccc}
\hline
\multirow{3}{*}{} & \multirow{3}{*}{\shortstack{Decoding \\ Condition}} & \multicolumn{4}{c}{WER (\%)} \\
\cmidrule(r){3-6} 
~ & ~ & \multicolumn{2}{c}{dev} &  \multicolumn{2}{c}{test} \\
\cmidrule(r){3-4} \cmidrule(r){5-6}
~ & ~ & clean &  other &  clean &  other  \\
\hline \hline
TT-baseline & bm5 & 3.0 & 7.7 & 3.2 & 7.6 \\
\hline
mask word & from TT score & 3.1 & 8.0 & 3.2 & 7.9 \\
mask token & from TT score & 3.0 & 7.8 & 3.2 & 7.7  \\
\multirow{2}{*}{+ sim. ins.} & from TT score  & 3.0 & 7.8 & 3.2 & 7.6  \\
 & from Reference & 2.6 & 6.7 & 2.7 & 6.6 \\
 
\hline
\end{tabular}
\label{tab:libriresults}
\end{table}

%% file: Tables/ms_decode.tex
\begin{table}[tp]
\caption{\small {Results of the proposed MS-decode algorithm on the Librispeech data. Adding SpecAugment and non-causal encoder are two strategies to further improve the performance. "correct 5-best" indicates correcting all TT outputs in the n-best list. 10 mask solutions are sampled for each best. ``oracle" represents the lower bound of WER for the current system.}}
\scriptsize
\vspace*{2mm}
\centering
\begin{tabular}{l l cccc}
\hline
\multirow{3}{*}{} & \multirow{3}{*}{\shortstack{Decoding \\ Condition}} & \multicolumn{4}{c}{WER (\%)} \\
\cmidrule(r){3-6} 
~ & ~ & \multicolumn{2}{c}{dev} &  \multicolumn{2}{c}{test} \\
\cmidrule(r){3-4} \cmidrule(r){5-6}
~ & ~ & clean &  other &  clean &  other  \\
\hline \hline
\multirow{2}{*}{TT-baseline} & bm5 & 3.0 & 7.7 & 3.2 & 7.6 \\
~ & bm5(oracle) & 1.9 & 5.7 & 1.9 & 5.6 \\
MLM-SC & from TT score & 3.0 & 7.8 & 3.2 & 7.6 \\
\hline
\multirow{4}{*}{+ MS-decode} & correct 1-best & 2.8 & 7.3 & 3.0 & 7.1  \\
  ~ & correct 1-best (oracle) & 2.7 & 6.8 & 2.8 & 6.6  \\
  ~ & correct 5-best & 2.6 & 6.7 & 2.8 & 6.7  \\
  ~ & correct 5-best (oracle) & 1.7 & 5.1 & 1.8 & 5.0  \\
\hline
 \multirow{2}{*}{+ SpecAug\cite{ParkCZCZCL19}} & correct 1-best & 2.8 & 7.0 & 2.9 & 7.0  \\

~ & correct 5-best & 2.5 & 6.5 & 2.7 & 6.6  \\
 \multirow{2}{*}{+ Non-Causal Enc.} & correct 1-best & 2.7 & 7.0 & 2.9 & 6.9  \\
 ~ & correct 5-best & 2.5 & 6.4 & 2.7 & 6.5  \\
 
\hline
\end{tabular}
\label{tab:ms_decode}
\end{table}

%% file: Tables/overall_results.tex
\begin{table}[tp]
\caption{\small {Overall Results on both the Librispeech and Aishell data. RTF: real time factor.}}
\scriptsize
\vspace*{2mm}
\centering
\begin{tabular}{c l cccc c}
\hline
\multirow{3}{*}{} & \multirow{3}{*}{\shortstack{Decoding \\ Condition}} & \multicolumn{2}{c}{WER (\%)} & \multicolumn{2}{c}{CER (\%)} & RTF \\
\cmidrule(r){3-6} \cmidrule(r){7-7}
~ & ~ & \multicolumn{2}{c}{libri-test} &  \multicolumn{2}{c}{aishell} & \ libri-test \\
\cmidrule(r){3-4} \cmidrule(r){5-6} \cmidrule(r){7-7}
~ & ~ & clean &  other &  dev &  test & other\\
\hline \hline
\multirow{4}{*}{\shortstack{Transformer \\-Transducer}} & bm5 & 3.2 & 7.6 & 6.7 & 7.3 & 0.1319 \\
~ & + rescore  & 2.9 & 7.0 & 6.4 & 6.9 & 0.1328 \\
~ & bm50 & 3.2 & 7.4 & 6.6 & 7.2 & 1.6183 \\
~ & + rescore & 2.9 & 6.7 & 6.3 & 6.8 & 1.6311 \\
\hline
\multirow{2}{*}{\parbox{1.5cm}{MLM-SC\\+MS decoding}} & correct 1-best & 3.0 & 7.1 & 5.8 & 6.3 & 0.1336 \\
~ & correct 5-best & 2.7 & 6.5 & 5.7 & 6.1 & 0.1387 \\
\hline
\end{tabular}
\label{tab:overall_results}
\end{table}

%% file: Tex/conclusion.tex
\section{Conclusion}
\label{sec:conclusion}

We propose to use audio-aware masked language model (MLM) as a spell correction model (MLM-SC). The advantage of using MLM-SC is that it is a non-autoregressive model, which is suitable for a second pass because MLM-SC causes low latency for the system. On the top of transformer-transducer (TT), we use MLM-SC to correct the first-pass results of TT. MLM-SC can achieve impressive improvements on the Aishell data measured by CER. However, there are no improvements when using MLM-SC on the Librispeech data measured by WER. To reduce this gap, we further propose a mask sample (MS) decoding algorithm. For each token with low confidence score, we do not directly mask the token. Instead, we sample between ``mask'' and ``not mask'', indicating that the token has a chance to be not masked. With multiple low confidence tokens, we can sample multiple mask solutions. They are fed into MLM-SC to be corrected. When we obtain multiple corrected utterances, we use a language model to select the best one as the final result. MLM-SC + MS-decode is evaluated on the Librispeech and Aishell1 data. Finally, the proposed methods achieve 7.0\% and 11.6\% relative error improvements compared to the TT-baseline with rescoring on the Librispeech and Aishell1 data, respectively.